\documentclass[10pt,letterpaper]{article}
\usepackage{opex3}
\usepackage{color}
\usepackage{graphicx}
\usepackage{bm}
\usepackage{float}
\usepackage{amsmath}
\usepackage{cite}

\newcommand{\ket}[1]{\vert #1 \rangle}

\begin{document}

\title{Dual-rail optical gradient echo memory}

\author{D.~B.~Higginbottom,$^{1,5}$ J.~Geng,$^{1,2,6}$ G.~T.~Campbell,$^1$ M.~Hosseini,$^1$ M.~T.~Cao,$^3$ B.~M.~Sparkes,$^1$ J.~Bernu,$^1$ N. P.~Robins,$^4$ P.~K.~Lam$^{1}$ and B.~C.~Buchler$^{1,7}$} 

\address{$^1$Centre for Quantum Computation and Communication Technology, Department of Quantum Science, The Australian National University, Canberra, ACT 0200, Australia\\
$^2$Department of Physics, State Key Laboratory of Precision Spectroscopy, East China Normal University, Shanghai 200062, P.~R.~China\\
$^3$Department of Applied Physics, School of Science, Xi'an Jiaotong University, Xi'an, Shaanxi, 710049, P.~R.~China\\
$^4$Quantum Sensors and Atomlaser Lab, Department of Quantum Science, Australian National University, Canberra, ACT 0200, Australia\\
$^5$daniel.higginbottom@anu.edu.au\\
$^6$geng@iqo.uni-hannover.de\\
$^7$ben.buchler@anu.edu.au}

\begin{abstract}
We introduce a scheme for the parallel storage of frequency separated signals in an optical memory and demonstrate that this dual-rail storage is a suitable memory for high fidelity frequency qubits. The two signals are stored simultaneously in the Zeeman-split Raman absorption lines of a cold atom ensemble using gradient echo memory techniques. Analysis of the split-Zeeman storage shows that the memory can be configured to preserve the relative amplitude and phase of the frequency separated signals. In an experimental demonstration dual-frequency pulses are recalled with $35$\% efficiency, $82$\% interference fringe visibility, and $6^\circ$ phase stability. The fidelity of the frequency-qubit memory is limited by frequency-dependent polarisation rotation and ambient magnetic field fluctuations, our analysis describes how these can be addressed in an alternative configuration.
\end{abstract}

\ocis{(020.0020) Atomic and molecular physics; (020.1670) Coherent optical effects; (210.0210) Optical data storage; (210.4680) Optical memories.} 



Optical quantum memories for the coherent and reversible transfer of quantum information between light and matter are an essential component of quantum communication and quantum computation proposals \cite{Lvovsky2009}.  Despite substantial challenges, quantum memories are now demonstrating efficiencies approaching 90$\%$ \cite{Hosseini2011, Hedges2010, Sparkes2013a}, storage times of many seconds \cite{Longdell2005a, Dudin2013, Heinze2013a}, bandwidths of hundreds of MHz \cite{Saglamyurek2011, Afzelius2010a} and large multi-modal capacities \cite{Saglamyurek2011, Hosseini2009, Hosseini2011}. 

The performance of a practical quantum memory is contingent on the qubit encoding that is used. Any linear, single-mode memory is capable of storing photon number qubits, however inefficiency and loss in the memory or other network elements will decrease output state fidelity. For this reason an alternate encoding is often used, such as the polarisation, orbital angular momentum, path, or arrival time (time-bin) of a single photon. Memories for polarisation qubits have been demonstrated using electromagnetically induced transparency (EIT) \cite{Cho2010, Riedl2012a, Kupchak2015}, atomic frequency comb (AFC)  \cite{Clausen2012} memories, and Raman absorption \cite{England2012}. EIT has also been used to store orbital angular momentum qubits with high fidelity \cite{Nicolas2014}. Temporally multimode memories, such as AFC \cite{Saglamyurek2011} and gradient echo memories (GEM) \cite{Hosseini2009}, are suitable for time-bin qubits.

Here we present work based on the three-level GEM protocol \cite{Alexander2006, Longdell2008, Hetet2008c} towards a quantum memory for frequency qubits \cite{Roslund2013,Humphreys2014}. We extend the GEM protocol to make use of Zeeman sub-levels to store and recall two frequency channels, this provides a basis to store frequency qubits with a high fidelity. These split Zeeman levels have previously been used to store matched pulses with EIT \cite{MacRae2008a}. In the three-level GEM scheme a strong optical coupling field and a weak signal field form an off-resonant Raman transition in an ensemble of $\Lambda$-type three-level atoms. By applying a linear atomic frequency gradient along the optical axis of the ensemble a signal pulse can be transferred into a coherent atomic spin wave excitation between the two ground states. Signals absorbed by the broadened ensemble can be re-emitted on demand by reversing the frequency gradient such that the atomic coherence rephases. Because gradient-echo memories can limit losses by operating with optical fields far detuned from the excited state, storage efficiencies can be made significantly higher than similar EIT memories. GEMs have been operated with 87$\%$ efficiency in warm atomic vapour  \cite{Hosseini2011,Hosseini2011a} and 80$\%$ efficiency in a cold atomic ensemble with longer storage time \cite{Sparkes2013}.

\begin{figure}[t]
\centering \includegraphics[width=0.8\columnwidth]{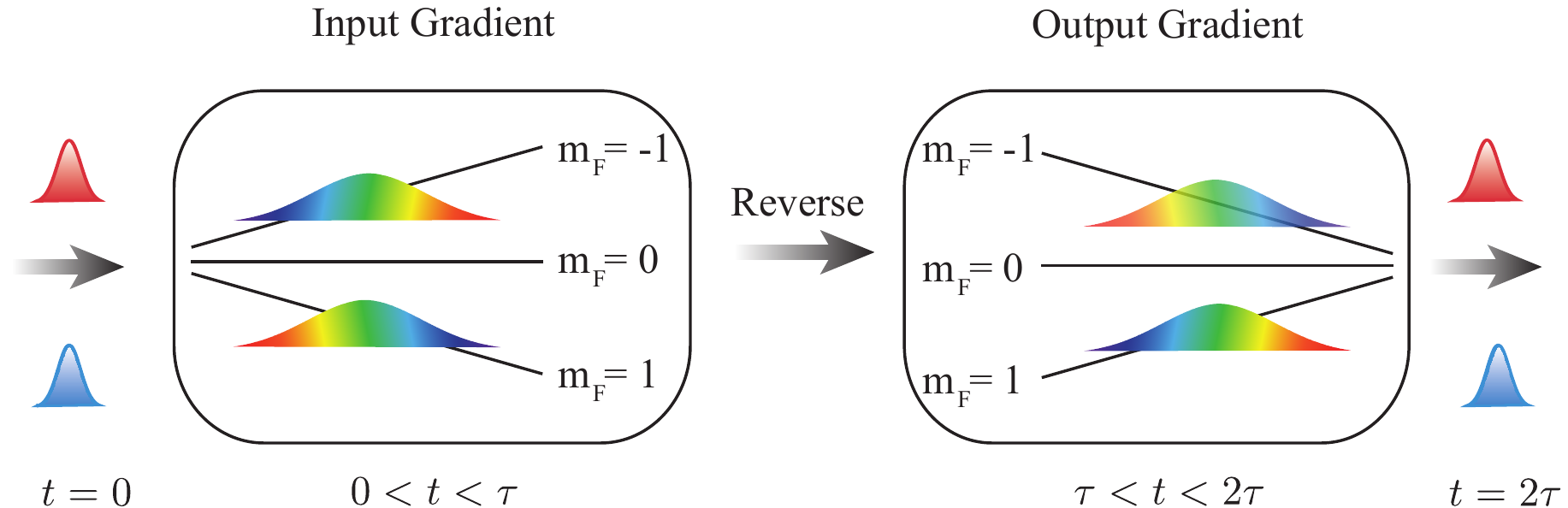} 
\caption{A simplified energy diagram of atomic levels $m_F = -1,0,1$ under constant uniform bias field $B_0$ and reversible gradient field $B_z$ illustrating the operation of the dual-rail gradient echo memory.}
\label{fig.simple_level}
\end{figure}

In our system the atomic frequencies are manipulated by applying a tailored magnetic field along the optical propagation axis. A bias magnetic field $B_0$ lifts the degeneracy of the Zeeman sub-levels by introducing a Zeeman shift given by $\Delta E = m_F g_F \mu_B B_0 /h$, where $m_F$ is the Zeeman level number, $g_F$ is the Land\'e factor for hyperfine state $F$, $\mu_B$ is the Bohr magneton and $h$ is the Planck constant. For the D1 line of $^{87}$Rb atoms the Land\'e factors are respectively $g_1=-1/2$ and $g_2=1/2$. A single coupling field detuned from the $F=2 \rightarrow F'=2$ transition will produce three Raman absorption lines for a weak signal field on the $F=1 \rightarrow F'=2$ transition, corresponding to signal field detunings of 0 (insensitive line) and $\pm\delta_0 $ relative to the unsplit Raman transition, where $\delta_0$ is given by 
 
\begin{equation} 
\delta_0 =  \frac{\mu_B}{h} B_0\simeq 1.4 \mathrm{ MHz/G} \times B_0. 
\end{equation} 

A reversible magnetic field gradient $B_z(z)$ is applied along the ensemble in addition to the uniform bias field $B_0$ to broaden the two sensitive lines around $\pm\delta_0$ for light storage using the GEM protocol, as illustrated in Fig~\ref{fig.simple_level}. Previous GEM experiments used only one of these two magnetically sensitive lines to store and recall pulses. In this paper we show that the sub-levels responsible for the two absorption lines can be utilised to store pulses of light at frequencies $\pm\delta_0$ in parallel such that the two lines form a memory for frequency encoded qubits. The same scheme readily extends to higher mode frequency states in atomic ensembles with high ground state angular momentum. For example, the four magnetically sensitive Zeeman levels of the $^{85}$Rb ground state (F=2) could be used to store four mode frequency states using the same approach that we demonstrate here with $^{87}$Rb.

 \begin{figure}[tb]
\centering \includegraphics[width=0.6 \columnwidth]{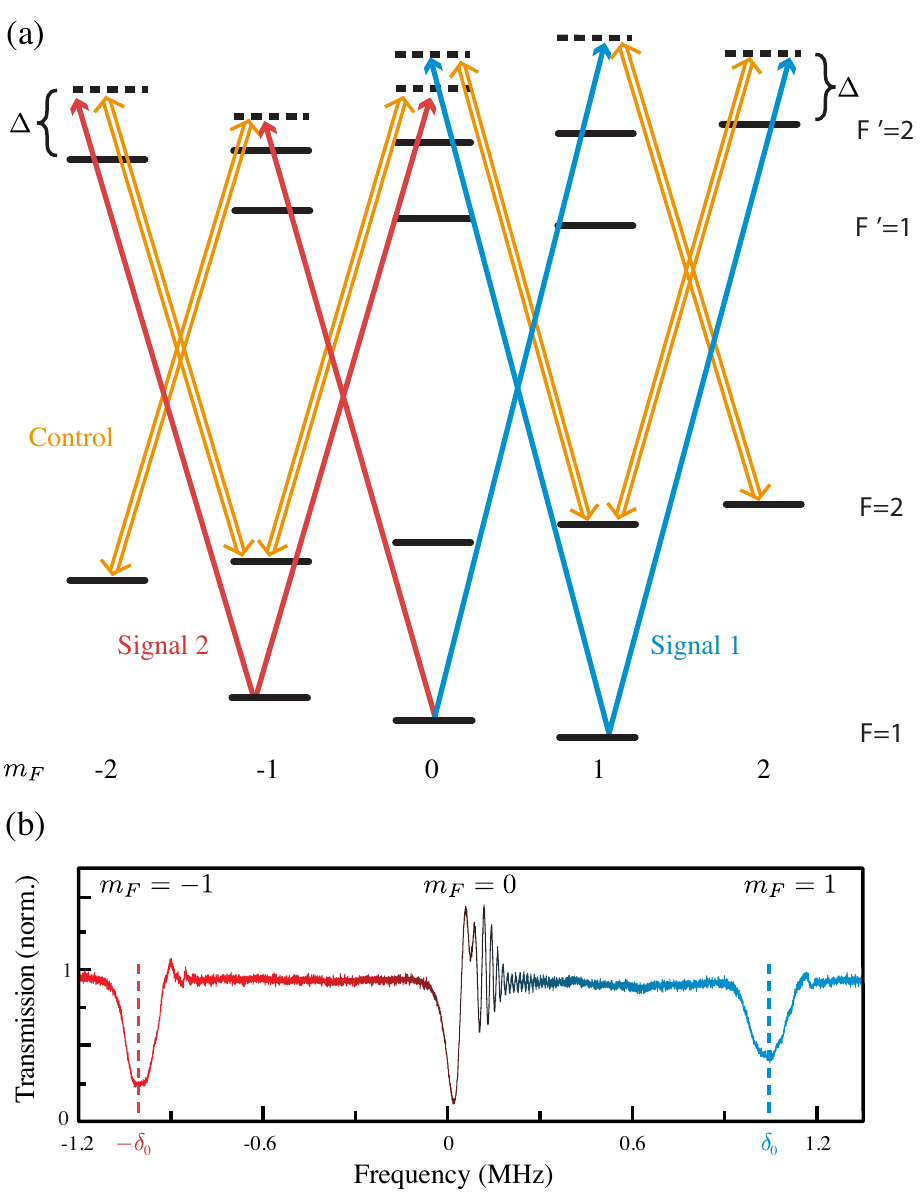} 
\caption{(a) Zeeman levels of the $^{87}$Rb D1 line split by a bias magnetic field $B_0$. The control and signal laser fields couple the two ground states F=1 and F=2 via multiple $\Lambda$-type Raman transitions. (b) Split Raman absorption lines for the dual-GEM scheme. Non-degenerate lines at $\pm \delta_0$ are produced by the bias magnetic field $B_0$ according to the level scheme above, and broadened by an additional field gradient $B_z$ that is applied to the atom ensemble for GEM. Free-induction decay can be seen in the centre line. Inhomogeneous broadening by the magnetic field suppresses free induction decay on the two magnetically sensitive lines.}
\label{fig.raman_lines}
\end{figure}

To achieve high fidelity frequency qubit storage in such a dual-rail GEM it is necessary to balance the effective optical depth of the two Raman lines shown in Fig.~\ref{fig.raman_lines}(b) and match the output polarisation of each frequency rail. The optical depth of each line depends on the dipole matrix elements associated with the transitions for the coupling and signal fields involved in the relevant two-photon transitions, as well as the polarisation of the two beams. In general, for an ensemble of atoms that has a population distributed uniformly across the Zeeman sub-levels, this results in unequal optical depths for the $\pm\delta_0$ Raman transitions. To balance the efficiencies of the two frequency rails we use linear polarisations for the coupling and signal fields which, by symmetry, ensures equal optical depths. In this configuration, however, frequency dependent polarisation rotation will produce output modes that are not matched.

Polarisation rotation occurs as a result of interference between coherences driven by the $\sigma^+$ and $\sigma^-$ components of each signal field. To examine the effect of this rotation, we consider below the level structure of the Rb$^{87}$ atoms, shown in Fig.~\ref{fig.raman_lines}(a). Off-resonant dispersion of $\sigma^+$ and $\sigma^-$ will further rotate the signal fields, but the relative Faraday rotation between signals 1 and 2 is negligible so long as their splitting, $\delta_0$, is much smaller than the detuning, $\Delta$.  

We assume that the population initially rests entirely in the $F=1$ ground state, evenly distributed between Zeeman sub-levels, and that there are no coherences within that manifold. These assumptions are valid provided that the probe field is weak and that the splitting, $\delta_0$, is large compared to the Raman linewidth. We proceed by treating each of the signals independently, starting with the signal 1. Referring to Fig.~\ref{fig.raman_lines}(a), signal 1 will produce a coherence, $\hat{\sigma_1}$, between the $\ket{F=1,m_F=1}$ and $\ket{F=2,m_F=1}$ states via three $\Lambda$ systems with excited states $\ket{F'=2,m_F=0}$, $\ket{F'=2,m_F=2}$ and $\ket{F'=1,m_F=0}$. This last $\Lambda$ system reduces the absorption asymmetry between $\sigma^+$ and $\sigma^-$ components of the signal. Consequently the $\sigma^+$/$\sigma^-$ absorption ratio approaches unity as the detuning $\Delta$ is increased.  

The $\Lambda$ system formed by $\ket{F=1,m_F=0}$, $\ket{F'=2,m_F=1}$ and $\ket{F=2,m_F=2}$ makes a small contribution to the memory process due to weak effective coupling strength. We neglect this contribution in the analysis below because the optical depth associated with this process is $\approx 19$ times smaller than the optical depth of the three $\Lambda$ schemes contributing to the $\ket{F=1,m_F=1} \rightarrow \ket{F=2,m_F=1}$ coherence. 

The equations of motion for signal 1 are

\begin{align} 
 \partial_t \hat{\sigma_1}  &= -(\gamma'_{0} + i \delta') \hat{\sigma_1} + i g^-_{1} \hat{\mathcal{E}}^-_1 + i g^+_{1} \hat{\mathcal{E}}^+_1 \label{Maxwell_Bloch:1} \\ 
   (\partial_t + c \partial_z) \hat{\mathcal{E}}^{\pm}_1 &= i g^{\pm}_{1} N \hat{\sigma_1},  \label{Maxwell_Bloch:2} 
 	\end{align} 
with effective coupling strengths 
	\begin{align} 
 	  g^+_{1} &\propto \frac{\Omega^+}{\Delta_{22}} \mu_{(1,1;2,2)}\mu_{(2,1;2,2)}\\ 
   g^-_{1} &\propto \frac{\Omega^-}{\Delta_{22}} \mu_{(1,1;2,0)}\mu_{(2,1;2,0)} + \frac{\Omega^-}{\Delta_{21}}\mu_{(1,1;1,0)}\mu_{(2,1;1,0)}), 
	\end{align} 
 	where $\mu_{(m,n;p,q)}$ is the dipole matrix element corresponding to the $\ket{F=m,m_F=n} \rightarrow \ket{F'=p,m_F=q}$ transition, $\Omega^{\pm}$ are the Rabi frequencies associated with the right and left circularly polarised components of the control field, $\Delta_{22} = \Delta$ is the control field detuning from the  $\ket{F=2} \rightarrow \ket{F'=2}$ transition and $\Delta_{21} = \Delta + 816.7$ MHz \cite{Steck2001} is the control field detuning from the $\ket{F=2} \rightarrow \ket{F'=1}$ transition. We can further simplify this system by introducing the signal 1 coupled mode $\hat{\mathcal{E}}^{\rm cp}_{1}$ 
 	\begin{align} 
 	  g^{\rm cp}_{1} \hat{\mathcal{E}}^{\rm cp}_{1} &\equiv (g^-_{1} \hat{\mathcal{E}}^-_1 + g^+_1 \hat{\mathcal{E}}^+_1) \\ 
 	  g^{\rm cp}_1 &\equiv \sqrt{\vert g_1^-\vert^2 + \vert g_1^+\vert^2}. 
 \end{align} 
	From Eqs. (\ref{Maxwell_Bloch:1}) and (\ref{Maxwell_Bloch:2}) we derive Eqs. of motion for the coupled mode
 			\begin{align} 
 			  \partial_t \hat{\sigma_1}  &= -(\gamma'_{0} + i \delta') \hat{\sigma_1} + i g^{\rm cp}_1 \hat{\mathcal{E}}^{\rm cp}_1 \label{Maxwell_Bloch:3} \\ 
 			  (\partial_t + c \partial_z) \hat{\mathcal{E}}^{\rm cp}_1 &= i g^{\rm cp}_1 N \hat{\sigma_1}.  \label{Maxwell_Bloch:4} 
 			\end{align} 

\begin{figure}[tb]
\centering \includegraphics[width=0.9 \columnwidth]{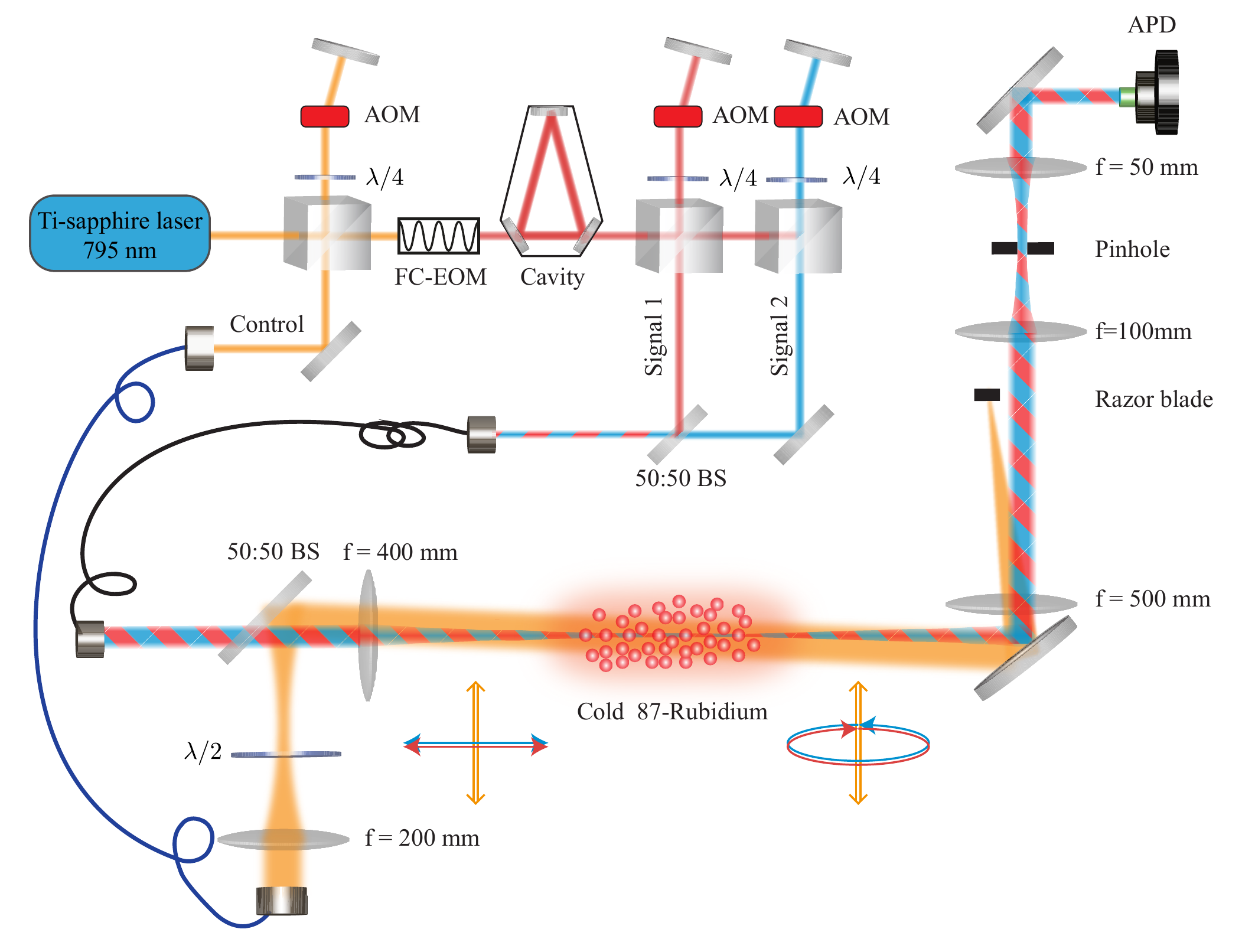} 
\caption{The experimental layout for dual-frequency GEM in an ensemble of cold $^{87}$Rb atoms. A root signal beam is produced from the control by ~6.8 GHz sideband generation (EOM) followed by frequency filtering (cavity). The beam is then split to allow independent gating and frequency control of signals 1 and 2 (AOMs). The two signal beams are recombined and sent to the atomic ensemble through an optical fibre. The $^{87}$Rb atoms are cooled and held in a magneto-optical trap setup similar to that in \cite{Sparkes2013, Geng2014}. The signal and control polarisations are indicated schematically before and after storage in the atom ensemble. The output signal polarisations  $\ket{P^{\rm cp}_1}$ and $\ket{P^{\rm cp}_2}$ (defined in the text) are elliptical with major/minor axes ratio 0.97/0.24. After passing through the atom ensemble the coupling field is spatially filtered using a pinhole and razor.}
\label{fig.setup}
\end{figure}

The component of the input perpendicular to the coupled mode polarisation is not stored in the memory. Given a horizontally polarised signal and vertically polarised control, $g^-_1 $ and $ g^+_1$ determine the signal 1 coupled mode polarisation $\ket{P^{\rm cp}_1} = 0.51\ket{L} + 0.86\ket{R}$. Analogous Eqs. exist for signal 2; by symmetry (assuming detunings $\Delta_{21}$ and $\Delta_{22}$ are approximately the same for signals 1 and 2, and given the same beam polarisations) the coupled mode polarisation for signal 2 is $\ket{P^{\rm cp}_2}  = 0.86\ket{L} + 0.51\ket{R}$. The input and output polarisations of both signals are indicated schematically in Fig~\ref{fig.setup}. The portion of the input orthogonal to the coupled mode is not stored in the memory, therefore the highest efficiency possible in this memory configuration is $\ket{H} \cdot \ket{P^{\rm cp}_1} = \ket{H} \cdot \ket{P^{\rm cp}_2} = 0.94$. The interference fringe visibility between the two output modes is limited to the coupled mode overlap $\ket{P^{\rm cp}_1} \cdot \ket{P^{\rm cp}_2} = 0.88$. 

To eliminate the polarisation rotation described above and improve fidelity the memory can be configured with circularly polarised signal and control beams. In this alternative configuration the memory fidelity is limited by recall efficiency asymmetry between the rails, which goes to zero in the high optical depth limit. With low optical depth one could still balance these efficiencies directly by preparing an ensemble population unevenly distributed between the Zeeman sub-levels of the ground state.

The experiment is carried out in a cold atomic cloud of ${}^{87}$Rb produced by a magneto-optical trap. The ensemble is initially prepared in the $F=1$ hyperfine ground state with a resonant optical depth of $\approx 300$ on the unsplit D$_1$ $F=1 \rightarrow F'=2$ transition. Details of the trapping and preparation of the ensemble can be found in \cite{Sparkes2013}. Because memory experiments require precise control of the magnetic field that is applied to the ensemble, the magnetic fields used for trapping must be switched off prior to performing the experiment. Eddy currents induced in the optical bench are allowed to die away for 1 ms after the trapping fields are removed before the memory experiment begins. The coupling field is turned on 400 $\mu$s before the writing stage to pump the atoms to the $F=1$ hyperfine ground state and remains on during the writing and reading stages.

The experimental setup is shown schematically in Fig~\ref{fig.setup}. The coupling and signal fields are derived from a Ti:Sapphire laser that is 200 MHz blue-detuned from the $^{87}$Rb D$_1$ $F=2 \rightarrow F'=2$ transition. The signal fields are produced by a fibre-coupled phase-modulator followed by a locked cavity. The modulator generates sidebands at $\approx \pm6.8$ GHz relative to the coupling field and the cavity is used to select only the correct sideband, which is then split into the two signal fields. All three fields pass through separate acousto-optic modulators to allow for fine frequency adjustment as well as gating and pulsing. The signal fields are combined with the same linear polarisation (horizontal), passed through an optical fibre and focused to match the size of the atomic cloud (beam waist = $200$ $\mu$m). The coupling field is polarised orthogonally to the signal fields and is collimated to a larger diameter of 7 mm so that it illuminates the entire ensemble uniformly. The coupling field propagates with a small angle relative to the signal beams and after passing through the cell it is extinguished by $\approx$ 45 dB using a razor blade and a pinhole. This spatial filtering incidentally reduces the signal field efficiency by 1 dB.

\begin{figure}[tb]
\centering \includegraphics[width=0.7 \columnwidth]{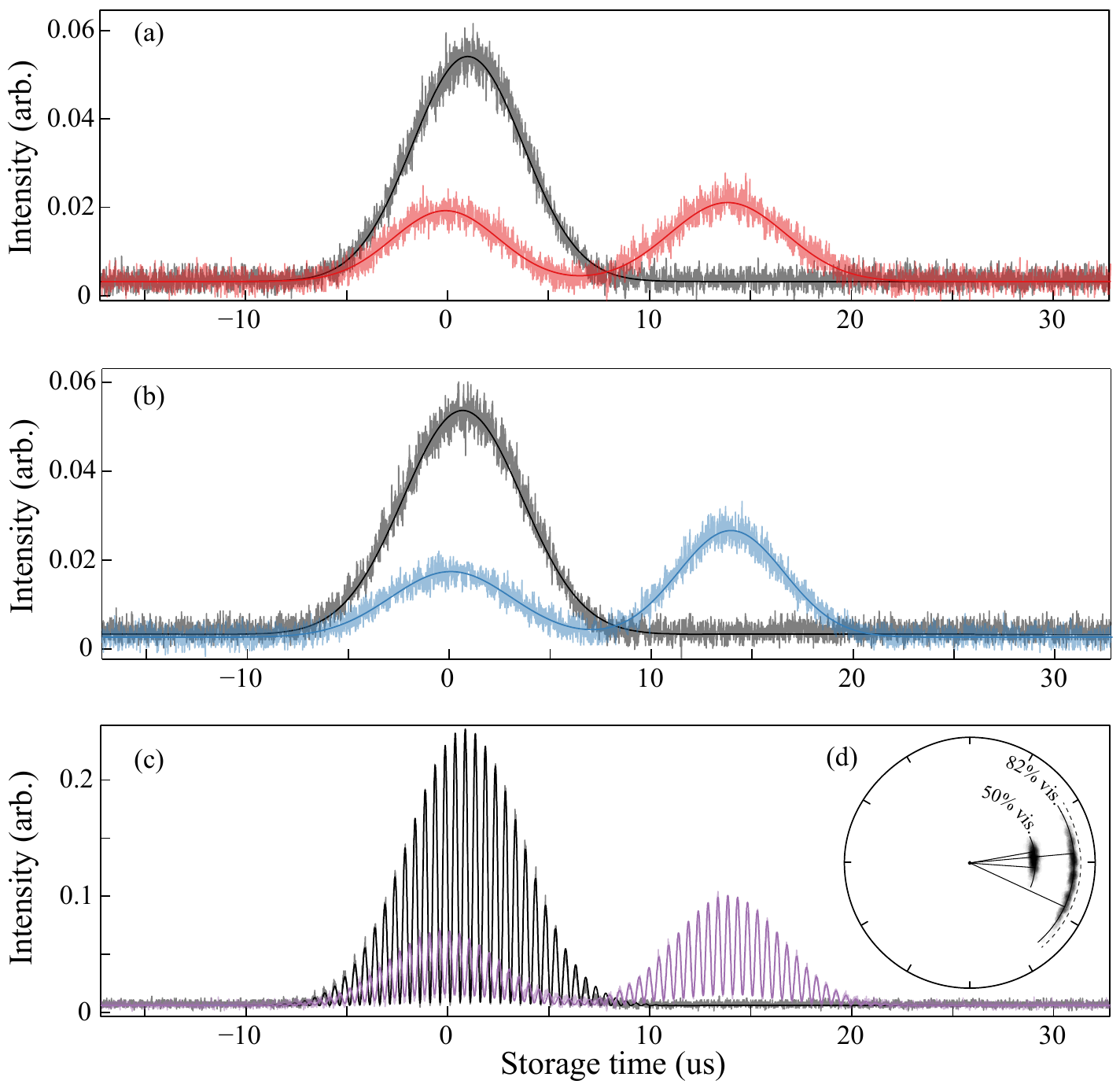}
\caption{Storage and recall of the two signal fields independently and simultaneously. (a) A typical input signal (black, recorded without the atomic ensemble) and echo pulse (red, also with light that has leaked through the memory during the write process) of signal 2 in isolation. (b) A typical input (black) and echo (blue) of signal 1 in isolation. (c) Temporally matched input pulses of both signals (black) and simultaneously recalled echoes (purple) demonstrating parallel frequency rail storage. Inset (d) shows the relative phase and fringe visibility between the two echo signals as an angle and radial displacement respectively. The first set of data has a phase stability of only $15^{\circ}$ and visibility of $82\%$. The black dotted arc is the visibility limit imposed by the polarisation distinguishability of the signal outputs. The phase stability of the second set was improved to $6^{\circ}$ by mains line triggering, but not optimised for high visibility interference}
\label{fig.echo}
\end{figure}

The results of the dual-rail storage and recall experiment are shown in Fig.~\ref{fig.echo}. Panels (a) and (b) show independent storage and recall of each of the signal fields in the absence of the other. Panel (c) shows storage of the two signal fields simultaneously. The input signal pulses have a Gaussian profile with 10 $\mu$s width. We record reference traces of the inputs (black) in the absence of the atomic ensemble. Traces showing storage and recall in the atom ensemble (red, blue, purple) are produced by reversing the magnetic field gradient across the MOT as per the GEM scheme described above such that the pulse is stored for approximately one pulse width. The polarisation of the control field is tuned to balance the efficiencies of the echoes to compensate for misalignment of the magnetic gradient relative to the optical axis and polarisation errors caused by the glass vacuum cell walls.

The recall efficiencies are $39\%$ and $32\%$ for signal fields 1 and 2 respectively. We demonstrate dual-rail storage by sending both signal pulses into the memory simultaneously with matched temporal profiles and amplitudes. The combined signals are recalled with 35$\%$ efficiency and the interference fringe visibility between the recalled pulses is $82\%$, compared to an expected visibility of $85\%$ given the output mode overlap derived earlier ($0.88$), the measured rail efficiencies ($0.32$,$0.39$) and the input temporal matching ($0.97$). The relative phase between the echo signals is measured by comparing the beat signal phase of the recalled pulses to earlier reference pulses that are not stored and not shown in Fig. \ref{fig.echo}. The storage process produces an overall shift in the phase difference between the two signals with a standard deviation of 15$^{\circ}$,  Fig.~\ref{fig.echo}(d). This variation is due to magnetic field background noise. By triggering the experimental sequence from the mains power supply, which improves the magnetic field stability, this phase noise was later reduced to 6$^{\circ}$, Fig.~\ref{fig.echo}(d). The output visibility in this second measurement was reduced by unrelated experimental considerations including exhaustion of our Rb source.

These results demonstrate that the dual-rail gradient echo memory can be used to store and recall two frequency separated signals in parallel while preserving the relative phase and amplitude of the optical states. This technique is therefore a suitable memory for frequency qubits with fidelity limited by the frequency-dependent polarisation rotation of two signals and the magnetic field stability of the atom ensemble. The alternative circular polarisation scheme described above is free of polarisation rotation. Instead the fidelity in this configuration is limited by unbalanced absorption of the two signals, which goes to zero as the optical depth increases and could be addressed even in a low optical depth ensemble by preparing asymmetric initial populations. Achievable improvements to the atom ensemble optical depth and ambient field isolation promise to improve both the efficiency and fidelity of such a frequency-qubit memory.

 \section*{Acknowledgements}
 This work is funded by the Australian Research Council Centre of Excellence Program (CE110001027). JG and MTC are supported by the Chinese Scholarship Council overseas scholarship.

\end{document}